\documentstyle[prb,aps,twocolumn,floats]{revtex}

\begin{document}

\twocolumn[\hsize\textwidth\columnwidth\hsize\csname@twocolumnfalse\endcsname
\draft
\title{Integrated Optical Devices Design by Genetic Algorithm}
\author{L. Sanchis, A. H\aa kansson, D. L\'opez-Zan\'on, J. Bravo-Abad
 and Jos\'e S\'anchez-Dehesa$^*$}
\address{Departamento de F\'{\i}sica Te\'orica de la Materia Condensada,
 Facultad de Ciencias (C-5),\\ Universidad Aut\'onoma de Madrid,
E-28049 Madrid, Spain.}

\date{\today}

\maketitle
\begin{abstract}

In this work we use multiple scattering in conjunction
with a genetic algorithm to reliably determine the optimized photonic-
crystal-based structure able to perform a specific optical task.
The genetic algorithm operates on a population of candidate structures to produce
 new candidates with better performance in an iterative process.
 The potential of this approach is illustrated by designing a spot size converter that
 has a very low F-number (F=0.47) and a conversion ratio
 of 11:1. Also, we have designed a coupler device that introduces
 the light from the optical fiber into a photonic-crystal-based waveguide
with a coupling efficiency over 87\% for a wavelength that can be tuned
 to 1.5 $\mu$m.
\end{abstract}
]

A new generation of optical devices is envisaged thanks to the properties of
photonic crystals (PC's).\cite{book} Though the recent advances in
three-dimensional PC's structures, in the last years much attention has been
focused on systems based on two-dimensional (2D) PC's because of their
easiness in the fabrication process. Thus, very compact optical devices and
circuits can be designed by introducing point and/or line defects. In order
to use such PC circuits in actual applications it is necessary to establish
a connection with an optical fiber. However, the core of the optical fiber
is about one order of magnitude larger than the PC-based waveguide.
Therefore, the design of an efficient (low loss) spot size converter is a
crucial goal in the field of PC; its solution will introduce the PCs devices
in the market place. In this regard, several groups\cite
{MMSpuler,AMekis,WKuang,PSanchis,DWPrather,HKosaka} have tackled this
problem by using different approaches. Most of them proposed tapered
waveguide structures\cite{MMSpuler,AMekis,WKuang,PSanchis}, or by using
reflective structures to focus the light into the waveguide.\cite{DWPrather}
A different approach consists of using the anisotropy of the PC's
equifrequency surfaces. \cite{HKosaka}

This letter introduces a method that is useful in determining the optimized
configuration of a 2D-PC structure capable of performing a requested optical
task with high efficiency . The method is illustrated by finding a spot size
converter (lens) that has a conversion ratio 11:1. In addition, the designed
PC structure that involves a spot-size converter in connection with a
PC-based waveguide it is presented. The insertion loss predicted for this
new structure is about 13\%, which is of the lowest reported by numerical
 simulations based on different coupling mechanisms.\cite
{MMSpuler,AMekis,WKuang,PSanchis,DWPrather} 

Our method is based on a binary-coded genetic algorithm (GA), an
optimization strategy inspired by Darwinian evolution\cite{JHHolland}. This
method has been applied to solve a wide variety of problems in different
fields like, for example, molecular geometry optimization \cite{DJWales},
material design\cite{GHJohanesson}, and artificial intelligence \cite
{HLipson}. In the field of optics, the GA has been employed in the synthesis
of Bragg gratings that conform to a particular spectrum,\cite{GCormier},
phase recovering from a fringe pattern\cite{FJCuevas}, and in designing
irregular lateral tapering.\cite{MMSpuler}

Although our proposal is general and applicable to any dimensionality, here
we analyze 2D-PCs for simplicity. We consider hexagonal lattice of Si
dielectric cylinders ($n_{Si}\approx $3.46 at 1.5 $\mu $m) in a background
of silica ($n_{SiO_{2}}$=1.45). We employ the multiple scattering theory
(MST) to study the diffraction effects of the TM modes (in-plane magnetic
field) on structures based on that symmetry. The MST has been successfully
applied by some of us in the analysis of metallic and dielectric clusters
based on 2D PCs \cite{TOchiai,JBravo}.

As the first step in the process, a structure made of a fixed number of
cylinders is chosen. The GA is then implemented with the possibility of
removing cylinders, but it cannot change neither their positions nor their
radius. A string of binary digits represents a possible structure ({\it %
individual)}. Each binary digit (1 means presence and 0 absence of a
cylinder at a given position) is called a {\it gene}. Each individual is
associated with a value, the fitness, which is strictly related to the PC
property that we are looking for. A crystal with a good-asked for property
will have a high fitness and vice versa. At start, a certain number of
individuals ({\it population}) is created by the GA that randomly determines
their genes. The size of population P depends on the complexity of the
problem, which is strictly related to the number of genes.

After the initiation, the GA selects two individuals at random, the one with
the highest fitness acts as a parent, the second parent is chosen in the
same way. Now, an uniform-crossover operator randomly exchanges the
chromosomes of the two parents with a probability C and, in this way, an
offspring is created. This is done until we have a new equal sized
population of offspring, i.e. a second generation. The selection and
crossover operators tend to enable the evolutionary process to move toward
promising regions of the search space. In a step further, a mutation
operator is introduced to prevent premature convergence to local maximum.
This operator changes the value of the genes at random with a probability $%
\mu $. A high value in the probability of mutation indicates that the GA
looks for better solutions at larger distances from a local maximum. The
three operators are repeatedly applied to the population, with the constrain
that the best-fitted individual is always copied into the next generation (%
{\it elitism}). As there is no way to know if the global maximum is reached,
this process is iterated until there is no improvement of the fitness value
of the best individual in the population. For further details the reader is
referred to books by Holland\cite{JHHolland}and Golberg.\cite{DEGoldberg}

The fitness parameter in our calculations is equal to the component of the
Poynting vector, $\overrightarrow{S}$, along the propagation direction of
the incident beam, which is evaluated in a selected position or in a set of
positions depending on the desired result. The Poynting vector is calculated
by the MST. The external radiation field is here a Gaussian beam in order to
demonstrate the spot-size conversion. The beam is formed as a sum of plane
waves weighted with a function dependent of the beam waist radius w$_{0}$
(see Ref. 15 and references therein). This beam represents the light at the
output of a fiber, its diameter 2w$_{0}$ being the diameter of the fiber
core (about 8-10 $\mu $m).

Let us deal with the design of the spot-size converter. We assume that a TM
polarized Gaussian beam (centered at the origin of our coordinate system)
propagates along the $x$-axis. This beam impinges a crystal consisting of a
hexagonal array of Si cylinders in silica, the $z$-axis being parallel to
the cylinder axis. The size of this crystal along the $y-$direction has to
be chosen slightly larger than the incident-beam-width to avoid flux
escaping at the lateral borders of the crystal, hence we have chosen a
crystal made of 26 rows of cylinders whose total length is about 3w$_{0}$
along the $y-$axis. Regarding its thickness along the $x$-axis, we used 13
columns of cylinders. The total number of cylinders is 318, which is
conditioned by our calculation resources. This structure constitutes the
lens-material (LM) of our integrated device. We consider a lattice constant $%
a_{LM}$ and radius $r=$0.294 $a_{LM}$. A band structure calculation of the
corresponding infinite crystal by means of a plane wave expansion method
predicts a forbidden gap for the TM-like modes in the range of frequencies
0.222-0.292 (in units of 2$\pi c/a_{LM}$). In order to have a low
reflectance device we choose a working frequency of 0.197; i.e. below the
first gap, where propagation is allowed in all directions. On the other
hand, the condition imposed to the GA in order to produce the spot-size
conversion is to maximize the $x$-component of the Poynting vector S$_{x}$
at one point (the focus) behind the cylinders. A full account of how the MST
is used to get this magnitude can be found in Ref. [15] and references
therein. The CPU time needed to get the fitness of one individual is of the
order of 6 s in a 2.8 GHz Pentium IV workstation. The coordinates of the
focal point ($x_{f},y_{f}$) are freely determined by us. Particularly, we
impose the focal point to be located on the symmetry axis of the system ($%
y_{f}=$0), and with a $x-$coordinate $x_{f}$=4.60 $\lambda _{0},$ where $%
\lambda _{0}$ is fixed by the working frequency ($\lambda _{0}\equiv a_{LM}$%
/0.197). Our results indicate that under this simple condition the
losses due to reflectance diminish and, at the same time, it considerably
reduces the waist of the propagated beam at the focus.

We realize that the global maximum must correspond to a symmetric structure,
which is obvious\ from the symmetry of the problem. Then, the GA is forced
to deal only with symmetric configurations. Therefore, the space search size
is reduced from 2$^{N}$ to 2$^{(N+n)/2}$ possible configurations, with N=318
and n=6, where N is the number of cylinders of the LM, and $n$ the number of
cylinders centered on the symmetry axis. In other words, from 5.3$\times $10$%
^{95}$ possible configurations to 5.8$\times $10$^{48}$. \ We have employed
a crossover operator with a probability C=0.4, which is maintained in the
rest of calculations, and a population P=400 individuals. The left panel in
Fig. 1 shows the resulting spot-size converter device and the pattern of the
corresponding electric field modulus. Two different scales are used to
measure distances, lattice parameter units (bottom and left) and wavelength
units (top and right). The
spot-size conversion can clearly be observed in the right panel of Fig. 1,
where $S_{x}$ is represented along the direction parallel to the $y-$axis at
the $x-$coordinate of the focal point (thin line), and for the incident beam
at the $y-$axis ($x=$0) (thick line) . A fit of $S_{x}$ ($x_{f}$, $y$) to a
Gaussian along the y-direction gives a beam waist radius of w$_{0}^{\prime }$%
=0.30 $\lambda _{0}$. This means that the spot-size converter has a
conversion ratio of 3.30 $\lambda _{0}$: 0.30 $\lambda _{0}\approx $ 11:1,
sufficient for classical-PC waveguide mode coupling requirements. Besides,
this lens-type structure shows another property that is difficult to achieve
by conventional lens design, its low F-number. According to Gaussian optics
F=(2$\pi )$w$_{0}^{\prime }$/4$\lambda _{0}$, which in our lens structure
takes a value of only 0.47. On the other hand, the power loss, which is
defined as the ratio of the transmitted power calculated inside the focus
waist (2w$_{0}^{\prime }$) to that of the incident beam, takes the value of
1.0 dB. In other terms, a 79\% of the incident power is squeezed and passes
through the focus. At this point let us remain that Kosaka et al. [7]
reported a photonic-crystal spot-size converter that reduces the spot-size
in the ratio 10:1, but no information is provided about the power passing
trough the focus.

Now, let us deal with the problem of designing an integrated device
involving the coupling of the light squeezed by the spot-size converted into
the PC waveguide. The parameters of the PC of what the waveguide is created,
which we name guide-material (GM), must be chosen carefully in order to
allow the coupling with a state localized in the waveguide. Firstly, the
photonic band structure has to have a full gap at the working frequency of
the LM and secondly, a guided mode of the crystal with a missing row of
cylinders, must exist at such frequency. Thus, we choose as the lattice
parameter of the GM, $a_{GM}$=1.5$a_{LM}$. On the other hand, we keep the
same orientation and cylinder radius as in the LM, and the PC waveguide is
created along the $\Gamma $K-direction in the hexagonal lattice. Figure 2
represents the TM-bands for the LM along $\Gamma $K as well as the projected
band structure of the GM along the missing row of cylinders. It is shown how
the working wavelength $\lambda _{0}$ of the lens matches a mode in the
guided band.

If one simply places the entry of the waveguide at the focal point of the
lens, one obtains a total insertion loss as high as 7.03 dB. This efficiency
is calculated as the ratio of the total power that is transmitted through
the waveguide to the incident power. The optimization process begins at this
point by defining a set of 52 cylinders, which are placed in front of the
entrance in order to facilitate the coupling to the waveguide \ mode, and
letting the GA operate over it. These cylinders have the same symmetry,
lattice parameter and orientation as the GM. The fitness was set equal to
the sum of $S_{x}$ calculated at 30 points located along a transversal
segment defined at the end of the waveguide; i.e., on the segment [-5.2 $%
a_{LM}$ , +5.2 $a_{LM}$]. Now, each individual has only 32 genes due to
symmetry reduction, and just 2$^{32}\simeq $ 4.3$\times $10$^{9}$ possible
combinations are available. The line (1) in Fig. 3 represents the fitness of
the best individual as a function of the number of generations.\ Each
generation contains a population P of 400 individuals. The structure with
the best fitness was achieved after 34 generations, the dashed horizontal
line defines the maximum value obtained. A resulting power loss of 1.07 dB
is obtained, which represents a substantial advance in comparison with the
total loss when no mouth-structure was considered.

In order to reduce the insertion loss further, we allow the GA to act over
the full structure made of the lens and the mouth. Now, its individual
contains 162+32 genes and, consequently, the space search size has been
increased up to 2$^{162+32}\simeq $2.5$\times $10$^{58}$. In this case we
consider that each generation contains a population of 600 individuals.
Figure 3 shows the result of running the GA on this structure, and Fig. 4
plots the optimized integrated device (the wave-guide coupler) together with
its electric field modulus pattern. The insertion loss predicted for this
structure is as low as 0.61 dB. This means that 87\% of the impinging light
passes through the waveguide and is detected at the output. In fact, this
value is underestimated since it does not include the light reflected at the
end of the waveguide by finite size effect. Therefore, the coupling
efficiency predicted by this new structure is comparable with the ones
reported in the literature. Thus, a waveguide-to-fiber coupling improvement
exceeding 2dB per converter is shown in Ref. [2]. References [3] and [4]
reported two different tapered couplers that numerically have over 90\%
power transmission. The simulations by P. Sanchis {\it et al}. [5] reported
that a transmission over 84\% can be achieved if defects are put on a planar
photonic crystal tapered waveguide. Finally, the J-coupler proposed by
Prather {\it et al}. [6] predicts a coupling efficiency of 91\%. In spite of
the large coupling efficiencies reported for these devices, in comparison
with the one here designed, the low compactness and integrability of these
devices are their main drawbacks when they are compared with the one designed
by evolutionary programming. 

Although our simulations involved the simplifying assumption of 2D PC's, it
should be noticed that such 2D-periodic crystals can be studied in actual 3D
crystals. For example, our recent simulations of 2D prism\cite{JBravo}
reproduces fairly well the behavior of the actual structures in the
microwave regime. We conjecture that similar results might also be
obtainable in the optical regime by using a PC-slab sandwiched between
multilayers films with a large gap. Also, the preceding discussion focused
on the TM modes of a structure based on
''dielectric-scatterers-on-background''. However, based on the general
method presented here similar devices based on ''holes-in-dielectric''
structures can also be obtained.

In summary, this work has shown that a genetic algorithm used in combination
with multiple scattering is able to design photonic crystal-based structures
with specific optical properties. By using this approach, we have demonstrated
the viability of fabrication of waveguide couplers with low losses and a high
spot-size conversion ratio. Besides, they can be integrated in the same wafer
 with other planar structures in a simple single-step lithographic process.
 It can be said that this method represents a substantial advance in order
 to reach the ultimate goal in optical devices design, that is molding the
 flow of light.

This work was supported in part by the Spanish CICYT (Project No.
MAT2000-1670-C04), CAM (Ref. 07N/0059/2002), and the EU under Contract No.
PHOBOS-IST-1999-19009. We also acknowledges the computing facilities
provided by the {\it Centro de Computaci\'{o}n Cient\'{\i}fica} at the UAM.

\newpage 
\begin{figure}[tbp]
\caption{(Left panel) Focusing effect produced by a spot-size converter
(lens) based on a 2D photonic crystal (black dots). The pattern of the
electric field modulus is represented in a wide spatial region. The lenght
scales are given in terms of the lattice parameter employed in the design of
the lens, $a_{LM}$, as wells as in terms of the working wavelength of the
lens $\protect\lambda_0$. (Right panel) The x-component of the Poynting
vector represented at the focal point ($x_f$=4.60 $\protect\lambda_0$). A
scale color is used, red (blue) color means maximum (minimum) electric field
modulus.}
\end{figure}

\begin{figure}[tbp]
\caption{(Left panel) Photonic band structure of the photonic crystal
employed as lens material (LM) along the $\Gamma $K direction (see inset).
(Right panel) Projected band structure of TM modes in the guide material
(GM). The grey regions represents the continuum of extended states in 
the photonic crystal. The white regions are the photonic band gaps.
 The waveguide is formed by removing one row of Si rods as shown in the inset.
The black line defines the guided modes inside the waveguide. The frequencies
in both panels are given in reduced units of the respective lattice parameters,
 $a_{LM}$ and $a_{GM}$. The horizontal dashed line defines the working 
frequency of optical devices under design.}
\end{figure}

\begin{figure}[tbp]
\caption{Running the genetic algorithm on the lens+waveguide integrated
structure. Line (1) defines the fitness of the best structure when the
optimization only acts over the cylinders on the mouth of the waveguide, the
dashed line defines the maximum fitness achieved. Line (2) represents the
corresponding result when the GA acts over the cylinders in the lens and
mouth simultaneously. Line (3) being the averaged fitness over the total
population in this case. The vertical lines separate the regions where
different values of mutation parameter $\protect\mu $ are employed. }
\end{figure}

\begin{figure}[tbp]
\caption{The optimized waveguided-coupler device (dots) obtained by a
genetic algorithm. The orange rectangles enclosed the cylinders where the GA
is applied. A scale color is used; red (blue) color defines the maximum
(minimum) electric field modulus.}
\end{figure}


\begin{references}
\bibitem[*]{byline}  Author to whom correspondence should be addressed:
jose.sanchezdehesa@uam.es

\bibitem{book}  J.D. Joannopoulos, R.D. Mead, and J.N. Winn, {\it Photonic
crystals} (Princeton University Press, Princeton, NJ, 1995).

\bibitem{MMSpuler}  M.M. Sp\"{u}ler {\it et al.}, J. Lighwave Technol.{\bf 16%
}, 1680 (1998).

\bibitem{AMekis}  A. Mekis and J.D. Joannopoulos, J. Lighwave Technolg. {\bf %
19}, 861 (2001)

\bibitem{WKuang}  W. Kuang {\it et al.}, Opt. Lett. {\bf 27}, 1605 (2002).

\bibitem{PSanchis}  P. Sanchis {\it et al.}, Electron.. Lett. {\bf 38}, 961
(2002).

\bibitem{DWPrather}  D.W. Prather {\it et al.}, Opt. Lett. {\bf 27}, 1601
(2002).

\bibitem{HKosaka}  H. Kosaka {\it et al.}, Appl. Phys. Lett. 76, 268 (2000).

\bibitem{JHHolland}  J.H. Holland, {\it Adaptation in Natural and Artificial
Systems} (The University of Michigan Press, Ann Arbor, 1975)

\bibitem{DJWales}  D.J. Wales and H.A. Scherage{\it , }Science {\bf 285},
1368 (1999).

\bibitem{GHJohanesson}  G.H. Johannesson {\it et al.}, Phys. Rev. Lett. {\bf %
88}, art. no. 255506 (2002).

\bibitem{HLipson}  H. Lipson and J.B. Pollack, Nature {\bf 406}, 974 (2000).

\bibitem{GCormier}  G. Cormier, R. Boudreau and S. Theriault, J. Opt. Soc.
Am. B{\bf 18}, 1771 (2001).

\bibitem{FJCuevas}  F.J. Cuevas, J.H. Sossa-Azuela, and M. Servin, Opt.
Commun. {\bf 203}, 213 (2002).

\bibitem{TOchiai}  T. Ochiai and J. S\'{a}nchez-Dehesa, Phys. Rev. B{\bf 65}%
, art. no. 245111 (2002).

\bibitem{JBravo}  J. Bravo-Abad, T. Ochiai and J. S\'{a}nchez-Dehesa, Phys.
Rev. B{\bf 67}, 115116 (2003).

\bibitem{DEGoldberg}  D. E. Goldberg, {\it Genetic Algorithms in Search,
Optimization and Learning} (Addison Wesley, Reading, MA, 1989).
\end{references}
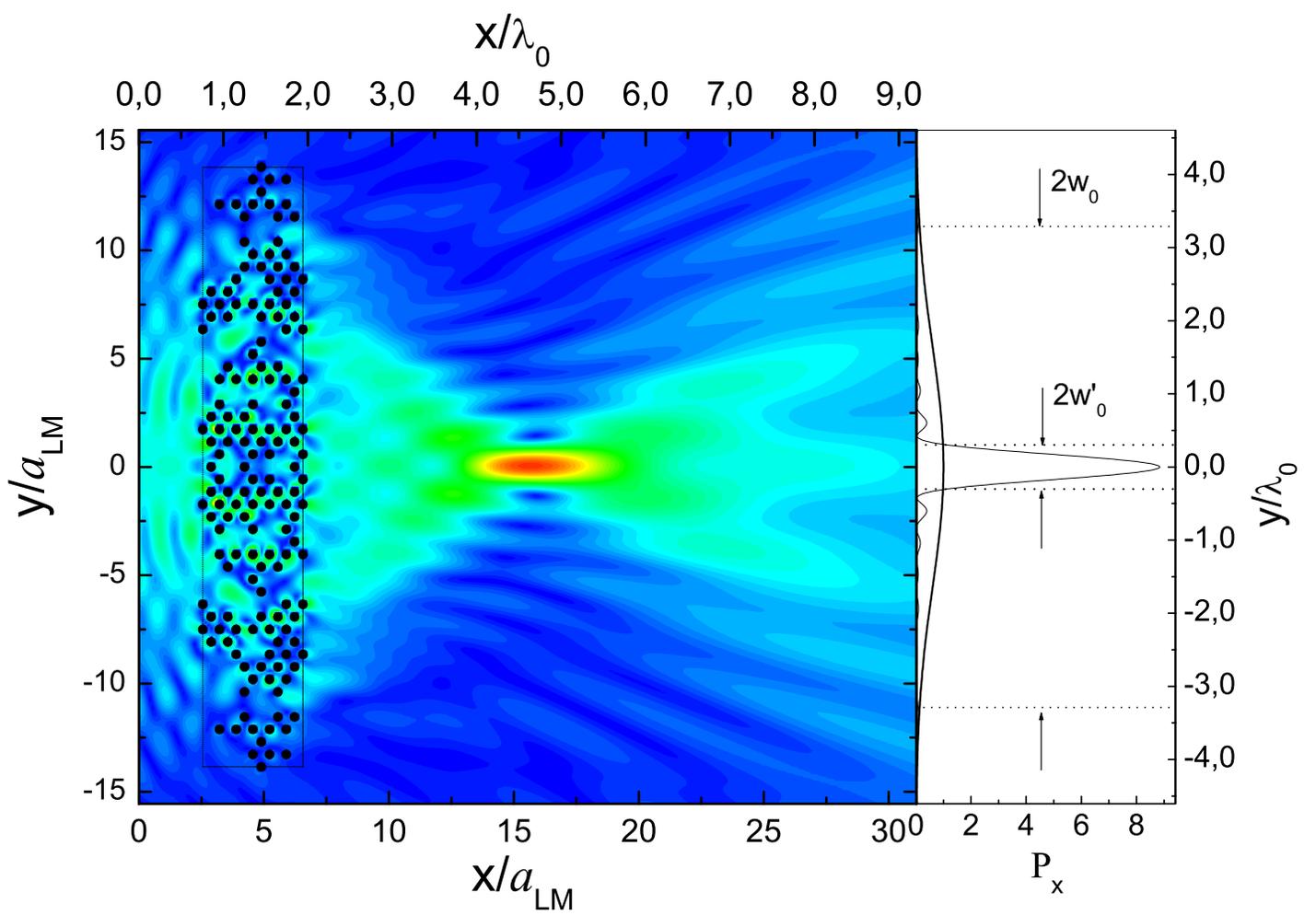
\end{document}